\documentstyle[preprint,aps]{revtex}
\preprint{USM-TH-107}
\begin{document}
\title{On the Interquark Potential calculation from Dirac Brackets}
\author{ Patricio Gaete \thanks{E-mail: pgaete@fis.utfsm.cl}}
\address{Departamento de F\'{\i}sica, Universidad T\'ecnica F.
Santa Mar\'{\i}a, Valpara\'{\i}so, Chile} \maketitle

\begin{abstract}
We obtain the binding energy of an infinitely heavy
quark-antiquark pair from Dirac brackets by computing the
expectation value of the pure QCD Hamiltonian. This procedure
exploits the rich structure of the dressing around static
fermions. Some subtle points related to exhibing explicitly the
interquark energy are considered.
\end{abstract}
\smallskip

PACS number(s): 12.38.Aw, 11.10.Ef

\section{INTRODUCTION}

It is well known that to gain a deeper insight into gauge theories
a proper study of the concepts of screening and confinement is of
considerable importance. In this connection the binding energy of
an infinitely heavy quark-antiquark pair represents a fundamental
concept in QCD which is expected to play an important role in the
understanding of quark confinement. In this respect we recall that
asymptotic freedom allows one to use perturbation theory for the
description of high energy phenomena. But within this framework,
we cannot explain low energy phenomena such as the permanent
confinement of quarks and gluons. The reason is that the infrared
divergences and gauge dependence make bound-state equations very
hard to approximate. We further note that the choice of the gauge
has a strong influence on the propagator. For instance, the
studies of Caracciolo et al. \cite{Caracc} were crucial in the
investigation of the ambiguity in the definition of the gluon
propagator.

In view of the above-mentioned difficulties much attention has
been recently devoted to formulations of QCD in which
gauge-invariant variables are explicitly constructed. By doing so
it has been possible to obtain a more direct physical insight into
the description of charged fields \cite{Lavelle,Prokhorov}. As a
result of this development the matter fields (quarks) are now
dressed by a cloud of gauge fields. Using this approach, the pure
QCD correction to the Coulomb potential in 3+1 dimensions has also
been derived more recently by Lavelle et al.  \cite{Lavelle2}. In
particular, it was showed that such correction contains a dominant
anti-screening contribution and another one which corresponds to
screening by physical gluons.

On the other hand, it is well known that gauge theories fall into
the class of constrained systems, where a systematic procedure for
quantizing such systems has been given by Dirac \cite{Dirac}. The
point we wish to emphasize, however, is that because of the
structure of the constraints the resulting equal-time commutation
relations involve in general the coupling constant. This raises
the question of how to recover from them the interquark potential
between charged fields following the conventional path via the
expectation value of the QCD Hamiltonian. This problem is
addressed in the present letter.

\section{INTERQUARK POTENTIAL}

Before computing explicitly the interquark potential, we shall
first present the Hamiltonian analysis for the Yang-Mills field
coupled to an external source $ J^0$. We start with the Lagrangian
\begin{equation}
{\cal L} =  - \frac{1}{4}Tr\left( {F_{\mu \nu } F^{\mu \nu } }
\right)-A_0^aJ^0 = - \frac{1}{4}F_{\mu \nu }^a F^{a\mu \nu }
-A_0^aJ^0. \label{rico1}
\end{equation}
Here $ A_\mu  \left( x \right) = A_\mu ^a \left( x \right)T^a $,
where $T^a$ is a hermitian representation of the semi-simple and
compact gauge group; and $ F_{\mu \nu }^a  = \partial _\mu  A_\nu
^a - \partial _\nu  A_\mu ^a  + gf^{abc} A_\mu ^b A_\nu ^c$, with
$f^{abc}$ the structure constants of the gauge group. The Dirac
procedure \cite{Dirac} as applied to (\ref{rico1}) is
straightforward. The canonical momenta are $ \Pi ^{a\mu } = -
F^{a0\mu }$, which results in the usual primary constraints $ \Pi
_0^a  = 0$, and $ \Pi ^{ai}  = F^{ai0}$. The canonical Hamiltonian
following from the above Lagrangian is:
\begin{equation}
H_c  = \int {d^3 x} \left( { - \frac{1}{2}\Pi _i^a \Pi ^{ai}  +
\Pi _i^a \partial ^i A ^{a0}  + \frac{1}{4}F_{ij}^a F^{aij}  -
gf^{abc} \Pi ^{ai} A ^{b0} A_i^c }+ A_0^a J^0 \right) .
\label{rico2}
\end{equation}
The persistence of the primary constraints leads to the following
secondary constraints
\begin{equation}
\Omega ^{a \left( 1 \right)} \left( x \right) = \partial _i \Pi
^{ai}  + gf^{abc} A ^{bi} \Pi _i^c - J^0  \approx 0 .
\label{secon}
\end{equation}
It is easy to check that there are no further constraints in the
theory, and that the above constraints are first class. The
corresponding total (first class) Hamiltonian that generates the
time evolution of the dynamical variables is given by
\begin{equation}
H = H_c  + \int {d^3 x\left( {c_0^a \left( x \right)\Pi _0^a\left(
x \right) + c_1^a \left( x \right)\Omega ^{a\left( 1 \right)}
\left( x \right)} \right)} , \label{rico3}
\end{equation}
where $c_0^a$ and $c_1^a$ are arbitrary functions. Since $ \Pi_0^a
\approx 0$ for all time and $ \dot{A}_0^a \left( x \right) =
\left[ {A_0^a \left( x \right),H} \right] = c_0^a \left( x
\right)$, which is completely arbitrary, we discard $ A_0^a \left(
x \right)$ and $ \Pi _0^a \left( x \right)$ because they add
nothing to the description of the system. The Hamiltonian then
takes the form
\begin{equation}
H = \int {d^3 } x\left( { - \frac{1}{2}\Pi _i^a \Pi _{ai}  +
\frac{1}{4}F_{ij}^a F^{aij}  + c^a \left( x \right)\left(
{\partial ^i \Pi _i^a  + gf^{abc} A^{bi} \Pi _i^c - J^0 } \right)}
\right) , \label{rico4}
\end{equation}
where $ c^a \left( x \right) = c_1^a \left( x \right) - A_0^a
\left( x \right)$.

Therefore, we have the first class constraints $ \Omega ^{a\left(
1 \right)} \left( x \right)$ , which appear at the secondary
level. Now the presence of the arbitrary quantities $c^a \left( x
\right)$ are undesirable since we have no way of giving them a
meaning in a quantum theory. To circumvent this trouble, we
introduce a supplementary condition on the vector potential such
that the full set of constraints becomes second class. A
particularly appealing and useful choice is given by
\begin{equation}
\Omega _a^{\left( 2 \right)} \left( x \right) = \int\limits_0^1
{d\lambda } \left( {x - \xi } \right)^k A_k^{\left( a \right)}
\left( {\xi  + \lambda \left( {x - \xi } \right)} \right) \approx
0   , \label{poinca1}
\end{equation}
where $\lambda(0\leq \lambda\leq1)$ is the parameter describing
the spacelike straight path $x^k=\xi^k+\lambda(x-\xi)^k$, on a
fixed time slice. This supplementary condition is the non-Abelian
generalization of the gauge condition discussed in
\cite{Gaete1,Gaete2}, which leads to the Poincar\'{e}
gauge\cite{Gaete3}. For simplicity we restrict our considerations
to $ \xi^k=0$. As a consequence, (\ref{poinca1}) becomes
\begin{equation}
\Omega _a^{(2)} \left( x \right) = \int\limits_0^1 {d\lambda } x^k
A_k^a \left( {\lambda x} \right) \approx 0 . \label{poinca2}
\end{equation}

Now we come to the calculation of the Dirac brackets. By following
the Dirac's procedure one arives at
\begin{equation}
\left\{ {A_i^a \left( x \right),A_b^j \left( y \right)} \right\}^
*   = 0 = \left\{ {\Pi _i^a \left( x \right),\Pi _b^j \left( y
\right)} \right\}^ * , \label{dirb1}
\end{equation}

\begin{equation}
\left\{ {A_i^a \left( x \right),\Pi ^{bj} \left( y \right)}
\right\}^ *   = \delta ^{ab} \delta _i^j \delta ^{(3)} \left( {x -
y} \right) - \int\limits_0^1 {d\lambda } \left( {\delta ^{ab}
\frac{\partial }{{\partial x^i }} - gf^{abc} A_i^c \left( x
\right)} \right)x^j \delta ^{(3)} \left( {\lambda x - y} \right) .
\label{dirb2}
\end{equation}
Note the presence of the last term on the right-hand side which
depends on $g$. In passing we note that similar Dirac brackets
were obtained independently in reference \cite{Poland}.

We are now equipped to compute the interaction energy between
pointlike sources in pure QCD, where a fermion is localized at the
origin $ {\bf 0}$ and an antifermion at $ {\bf y}$. In order to
accomplish this purpose, we will calculate the expectation value
of the energy operator $ H$ in the physical state $
|\Omega\rangle$, which we will denote by ${ \langle H \rangle_
\Omega}$ . From our above discussion, we see that ${ \langle H
\rangle_ \Omega}$ reads
\begin{equation}
\left\langle H \right\rangle _\Omega   = \left\langle \Omega
\right|\int {d^3 x} \left( { - \frac{1}{2}\Pi _i^a \Pi ^{ia}  +
\frac{1}{4}F_{ij}^a F^{aij} } \right)\left| \Omega  \right\rangle.
\label{sort1}
\end{equation}
Since the fermions are taken to be infinitely massive (static),
this can be further simplified as
\begin{equation}
\left\langle H \right\rangle _\Omega   = \left\langle \Omega
\right|\int {d^3 x} \left( { - \frac{1}{2}\Pi _i^a \left( x
\right)\Pi ^{ia} \left( x \right)} \right)\left| \Omega
\right\rangle . \label{sort2}
\end{equation}
Let us also mention here that, as was first established by Dirac
\cite{Dirac2}, the physical states $|\Omega\rangle$ correspond to
the gauge invariant ones. It is helpful to recall at this stage
that in the Abelian case $|\Omega\rangle$ may be written as
\cite{Gaete1}
\begin{equation}
\left| \Omega  \right\rangle  \equiv \left|\overline \Psi \left(
{\bf y} \right)\Psi \left( {\bf 0} \right)   \right\rangle =
\overline \psi \left( {\bf y} \right)\exp \left(
{ig\int\limits_{\bf 0}^{\bf y} {dz^i A_i \left( z \right)} }
\right)\psi \left( {\bf 0} \right)\left| 0 \right\rangle ,
\label{sort3}
\end{equation}
where $|0\rangle$ is the physical vacuum state and the line
integral appearing in the above expression is along a spacelike
path starting at $\bf 0$ and ending at $\bf y$, on a fixed time
slice. It should be clear from this discussion that the strings
between fermions have been introduced in order to have a
gauge-invariant function $ \left| \Omega \right\rangle $.
According to this viewpoint the fermion fields are now dressed by
a cloud of gauge fields.

The expression (\ref{sort3}) may be extended on account of the
fact that we have non-Abelian fields. Accordingly, we can write a
state which has a fermion at $ \bf 0$ and an antifermion at $\bf
y$ as
\begin{equation}
\left| \Omega  \right\rangle  = \overline \psi \left( {\bf y}
\right)U({\bf y},{\bf 0})\psi \left( {\bf 0} \right)\left| 0
\right\rangle, \label{sort4}
\end{equation}
where
\begin{equation}
U({\bf y},{\bf 0})\equiv P \exp \left( {ig\int\limits_{\bf 0}^{\bf
y} dz^i  A_i^a \left( z \right)T^a  } \right)  . \label{sort4b}
\end{equation}
As before, the line integral is along a spacelike path on a fixed
time slice, $P$ is the path-ordering prescription and $|0\rangle$
is the physical vacuum state.

This last point gives us an opportunity to compare our work with
the standard Wilson loop procedure \cite{Peskin}, to make sure
that the known results are recovered from the general expression
(\ref{sort4}) in the weak coupling limit. In effect, due to
asymptotic freedom, the short distance behavior of the interquark
potential is determined by perturbation theory. According to this,
at weak coupling, one can expand
\begin{equation}
U\left( {{\bf y},{\bf 0}} \right) \equiv P\exp \left(
{ig\int\limits_{\bf 0}^{\bf y} {dz^i } A_i^a \left( z \right)T^a }
\right) = P\left( {1 + ig\int\limits_{\bf 0}^{\bf y} {dz^i } A_i^a
\left( z \right)T^a + ...} \right) . \label{sort5}
\end{equation}
This implies that, at lowest order in $g$, the non-Abelian
generalization of the dressing framework is the same as in the
Abelian theory. Thus, at short distances, one should expect to
obtain the known Coulomb potential in addition to a correction of
order $g^4$, as we will now show.

With this in view, our next task is the computation of the
expectation value of $H$ in the physical state $|\Omega\rangle$
given by the expression (\ref{sort2}). From the above Hamiltonian
analysis one distinguishes here an Abelian part (proportional to
$C_F$) and a non-Abelian part (proportional to the combination
$C_FC_A$). We first consider the Abelian part which is identical
to the $QED$ case. To do this, we shall begin by observing that
\begin{equation}
\Pi _i^a \left( {\bf x} \right)\left| \Omega \right\rangle =
\overline \psi  \left( {\bf y} \right)U({\bf y},{\bf 0})\psi
\left( {\bf 0} \right)\Pi _i^a \left( {\bf x} \right)\left| 0
\right\rangle  + gT^a \int\limits_{\bf 0}^{\bf y} {dz_i } \delta
\left( {{\bf x} - {\bf z}} \right)\left| \Omega \right\rangle.
\label{RJ}
\end{equation}
Using this in (\ref{sort2}) we then evaluate the interaction
energy in the presence of the static charges
\begin{equation}
V^{g^2 }  = \frac{1}{2}g^2 trT^a T^a \int\limits_{\bf 0}^{\bf y}
{dz^i } \int\limits_{\bf 0}^{\bf y} {dz_i^\prime  } \delta \left(
{{\bf z} - {\bf z}^\prime  } \right), \label{SP}
\end{equation}
remembering that the integrals over $z^i$ and $z_i^\prime$ are
zero except on the contour of integration. Writing the purely
group theoretic factor $trT^aT^a=C_F$, the expression (\ref{SP})
leads to
\begin{equation}
V^{g^2}(L) =   \frac{1}{2}g^2 C_F k L, \label{energy2}
\end{equation}
after subtracting the term ${\langle H
\rangle}_0=\langle0|H|0\rangle$, where $|{\bf y}|\equiv L$ and $k
= \delta^{(2)}(0)$. This calculation shows that special care is
required in order to clarify the appearance of this peculiar
result. It may be remarked, however, that the origin of the
divergence is quite clear, so that it is possible to extract the
Coulomb potential from the infinite contribution. Notice that the
origin of the divergent factor $k$ is due to the fact that the
thickness of the string is nonvanishing only on the contour of
integration. It is worth stressing at this stage that a more
careful examination of the term $ \ \frac{{g^2 }}{2}\int {d^3
x\left( {\int\limits_{\bf 0 }^{\bf y} {dz_i \delta ^{(3)} \left(
{x - z} \right)} } \right)} ^2$ reproduces exactly the expected
Coulomb interaction between charges after subtracting the
self-energy term, as was discussed in \cite{Gaete1}. Having made
this observation, we write immediately the standard result for the
potential to order $g^2$, that is,
\begin{equation}
V^{g^2}(L) = - \frac{1}{{4\pi }}g^2 C_F \frac{1}{L}.
\label{energy3}
\end{equation}
Let us also mention here that if we had considered a modified form
for the supplementary condition (\ref{poinca1}), which is
equivalent to the Coulomb gauge \cite{Gaete1}, the result for the
potential would have been the same.

We now turn our attention to the non-Abelian part. From the
expressions (\ref{dirb2}) and (\ref{sort2}), we see that the $g^4$
contribution may be written in the form
\begin{equation}
V^{g^4 }  = \int {d^3 } x\left\langle 0 \right|\left( {I^i }
\right)^2 \left| 0 \right\rangle , \label{newcont}
\end{equation}
where
\begin{equation}
I^i  = g^2 f^{bac} T^b \int\limits_{\bf 0}^{\bf y} {dz^k }
\int\limits_0^1 {d\lambda } A_k^c \left( z \right)z^i \delta
\left( {{\bf x} - \lambda {\bf z}} \right) . \label{newcont2}
\end{equation}
The expression (\ref{newcont2}) may then be further manipulated as
described in \cite{Gaete1}. Thus, if we use spherical coordinates,
we find that
\begin{equation}
I^i  = g^2 f^{bac} T^b \frac{{{\bf z}^i }}{{|{\bf
z}|}}\frac{1}{{|{\bf x}|^2 }}\int\limits_{\bf 0}^{\bf y} {dz^k }
A_k^c(z) \sum\limits_{lm} {Y_{lm}^
* } \left( {\theta ^\prime  ,\varphi ^\prime  } \right)Y_{lm} \left(
{\theta ,\varphi } \right). \label{newcont3}
\end{equation}
Now, by employing (\ref{newcont3}) we can reduce (\ref{newcont})
to
\begin{equation}
V^{g^4 }  = \frac{1}{2}g^4 tr \int {d^3 x} \left\langle 0
\right|\left( {f^{bac} T^b \frac{{{\bf z}^i }}{{|{\bf
z}|}}\frac{1}{{|{\bf x}|^2 }}\int\limits_{\bf 0}^{\bf y} {dz^k }
A_k^c (z) \sum\limits_{lm} {Y_{lm}^ *  } \left( {\theta ^\prime
,\varphi ^\prime  } \right)Y_{lm} \left( {\theta ,\varphi }
\right)} \right)^2 \left| 0 \right\rangle , \label{newcont4}
\end{equation}
which, by introducing the integration variable ${\bf r}={\bf x}$
and using usual properties for the spherical harmonics, may be
rewritten as
\begin{equation}
V^{g^4 } \left( L \right) =   \frac{1}{2}g^4 C_A C_F \left( { -
\frac{1}{L}} \right) \int\limits_{\bf 0}^{\bf y} {dz^i }
\int\limits_{\bf 0}^{\bf y} {dz^{\prime j} } D_{ij}(z,z ^\prime) .
\label{newcont5}
\end{equation}
Here $D_{ij}(z,z ^\prime)$ is the propagator, which is diagonal in
colour and taken in an arbitrary gauge. Thus, in order to carry
out this calculation, we choose, for example, $D_{ij}(z,z
^\prime)$ in the Feynman gauge. Hence expression (\ref{newcont5})
reduces to
\begin{equation}
V^{g^4 } \left( L \right) = - \frac{1}{{8\pi ^2 }} g^4 C_A C_F
\left( { - \frac{1}{L}} \right)\int\limits_{\bf 0}^{\bf y} {d{\bf
z}} \int\limits_{\bf 0}^{\bf y} {d{\bf z}^\prime }
\frac{1}{{\left( {z - z^\prime } \right)^2 }} . \label{newcont6}
\end{equation}
This allows us to derive the $g^4$ contribution
\begin{equation}
V^{g^4 } \left( L \right) = -g^4 \frac{1}{{4\pi ^2 }} C_A C_F
\frac{1}{L}\log \left( {\mu L} \right), \label{newcont7}
\end{equation}
where $\mu~$ is a cutoff. By putting together Eqs.(\ref{energy3})
and (\ref{newcont7}), we obtain for the total interquark potential
\begin{equation}
V\left( L \right) =  - g^2 C_F \frac{1}{{4 \pi L }}\left( 1 +
\frac{g ^2}{{\pi }}C_A \log \left( {\mu L} \right) \right).
\label{newcont8}
\end{equation}

In this way one obtains the known heavy interquark potential at
lowest order in $g$ \cite{Lavelle2}. However, the central
difference between the above analysis and that of Ref.
\cite{Lavelle2} rests in the fact that the potential
(\ref{newcont8}) is directly recovered from the constraints
structure of the theory we have discussed. In this context, the
present investigation complements the discussion done in
Ref.\cite{Lavelle2}, as well as it reveals the general viability
of our analysis. Thus it seems a challenging job to extend the
scope of applicability of the above analysis. We expect to report
on progress along these lines soon.

\section{ACKNOWLEDGMENTS}

I would like to thank G. Cvetic for helpful comments on the
manuscript. Work supported in part by Fondecyt (Chile) grant
1000710, and by a C\'atedra Presidencial (Chile). I would also
like to thank I. Schmidt for his support.

\end{document}